\documentclass[08pt,twocolumn,preprintnumbers,amsmath,amssymb,nofootinbib,superscriptaddress,prd]{revtex4-1}
\usepackage{amsmath,amssymb,graphicx}
\usepackage{float}
\usepackage{latexsym} 
\usepackage{amsmath,amsthm}
\usepackage{ifpdf}
\usepackage{epstopdf}
\usepackage{color}
\usepackage{dcolumn}% Align table columns on decimal point
\usepackage{bm}% bold math
\usepackage{braket}
\usepackage[symbol]{footmisc}
\usepackage{graphicx}
\DeclareMathOperator{\sech}{sech}
\DeclareMathOperator{\arcsech}{arcsech}

\newcommand{\hatF}{\hat{\phi}}%

  \newcommand{\D}{\,\mathrm{d} }
\begin{document}
\def \beq{\begin{equation}}
\def \eeq{\end{equation}}
\def \bea{\begin{eqnarray}}
\def \eea{\end{eqnarray}}
\def \bes{\begin{split}}
\def \ees{\end{split}}
\def \besu{\begin{subequations}}
\def \esu{\end{subequations}}
\def \bea{\begin{align}}
\def \eal{\end{align}}
\def \bem{\begin{displaymath}}
\def \eem{\end{displaymath}}
\def \P{\Psi}
\def \Pd{|\Psi(\boldsymbol{r})|}
\def \Pds{|\Psi^{\ast}(\boldsymbol{r})|}
\def \Po{\overline{\Psi}}
\def \bs{\boldsymbol}
\def \dert{\frac{d}{dt}}
\def \k{\ket}
\def \br{\bra}
\def \bm{\hat b^-_{\Omega}}
\def \bp{\hat b^+_{\Omega}}
\def \am{\hat a^-_{\omega}}
\def \ap{\hat a^+_{\omega}}
\def \pau {\partial_u}
\def \pav{\partial_v}
\def \paut{\partial_{\tilde u}}
\def \pavt{\partial_{\tilde v}}
\def \a{\alpha_{\Omega \omega}}
\def \b{\beta_{\Omega \omega}}

\title{Sine-Gordon soliton as a model for Haw\-king radiation of moving black holes and quantum soliton evaporation}
\author{Leone Di Mauro Villari}
\affiliation{Institute of Photonics and Quantum Sciences, Heriot-Watt University, Edinburgh EH14 4AS, (UK)}
\affiliation{Institute for Complex Systems, National Research Council, (ISC-CNR), Via dei Taurini 19, 00185, Rome (IT)}
\author{Giulia Marcucci}
\thanks{Corresponding author. Email: marcucci.giulia@gmail.com}
\affiliation{University of Rome La Sapienza, Department of Physics, Piazzale Aldo Moro 5 00185, Rome, Italy}
\affiliation{Institute for Complex Systems, National Research Council, (ISC-CNR), Via dei Taurini 19, 00185, Rome (IT)}
\author{Maria Chiara Braidotti}
\affiliation{University of L'Aquila, Department of Physical and Chemical Sciences, Via Veneto 10, I-67010 L'Aquila (IT) }
\affiliation{Institute for Complex Systems, National Research Council, (ISC-CNR), Via dei Taurini 19, 00185, Rome (IT)}
\author{Claudio Conti}
\affiliation{Institute for Complex Systems, National Research Council, (ISC-CNR), Via dei Taurini 19, 00185, Rome (IT)}
\affiliation{University of Rome La Sapienza, Department of Physics, Piazzale Aldo Moro 5 00185, Rome, Italy}

\begin{abstract}
The intriguing connection between black holes' evaporation and physics of solitons is opening novel roads to finding observable phenomena. It is known from the inverse scattering transform that velocity is a fundamental parameter in solitons theory.  Taking this into account, the study of Haw\-king radiation by a moving soliton gets a growing relevance. However, a theoretical context for the description of this phenomenon is still lacking. Here, we adopt a soliton geometrization technique to study the quantum emission of a moving soliton in a one-dimensional model. Representing a black hole by the one soliton solution of the sine-Gordon equation, we consider Haw\-king emission spectra of a quantized massless scalar field on the soliton-induced metric. We study the relation between the soliton velocity and the black hole temperature. Our results address a new scenario in the detection of new physics in the quantum gra\-vi\-ty panorama.
\end{abstract}

%Use showkeys class option if keyword

\maketitle
\section{Introduction}
During the last ten years, analogue gra\-vi\-ty systems have attracted major interest in the scientific community~\cite{Liberati}. These models aim at providing valuable scenarios to test inaccessible features of quantum gra\-vi\-ty, as the Haw\-king radiation emission by black holes (BHs)~\cite{hawking74}. Furthermore, the recent observation of gra\-vi\-ta\-tio\-nal waves (GWs) emitted by colliding BHs~\cite{GW1,GW2} shaded new light and opened unexplored roads towards the search for quantum effects in gra\-vi\-ty~\cite{mwi}, as the Haw\-king's BH evaporation~\cite{sakalli2016,*sakalli2015,*sakalli2015bis,*sakalli2016bis}.
Indeed, quantum BH emission might be observed by the concomitant monitoring of the BH collisions by gra\-vi\-ta\-tio\-nal and electromagnetic antennas.
However, the collision process changes the original Haw\-king's framework.%% \st{, and calls for including the BH motion in the theoretical analysis}.
%% \st{At the collision, the BH velocity drops from $0.5c$ ($c$ being the vacuum light velocity) to zero; we do not know the way this dramatic change affects the BH temperature and the corresponding Haw\-king's spectrum.} \st{But, following the original Unruh argument %\cite{Unruh1976},
%% one can imagine a sort of chirp in the Haw\-king spectral emission, which may be eventually observable and extracted within measured electromagnetic signals.}
%% \st{At the moment, we still miss simple theoretical frameworks for treating these problems.}

Originally, Haw\-king considered quantum fields in a stationary BH background, the Schwarzschild metric, and discovered that BHs emit thermal radiation and evaporate.  %\st{However, the Schwarzschild BH is a particular solution of the Einstein field equations, so one could argue if it is the only possible choice.}
His paper appeared exactly one year after a trailblazing article by Ablowitz, Kaup,  Newell and Segur (AKNS), that cast new light on nonlinear waves by e\-sta\-bli\-shing the general method to solve classes of nonlinear field equations~\cite{ablowitz73,ablowitz73a}. Surpri\-singly, AKNS classes generate a metric and define an event horizon (EH). Indeed, it is known in the field of the nonlinear waves that integrable systems, which can be solved exactly by the inverse scattering transform (IST), describe a Riemannian surface with constant negative curvature~\cite{Bullough97,sasaky79}.

Recently, Haw\-king radiation analogues from solitons were considered in a huge variety of physical contexts, including light~\cite{lp,s,bl,tc,villari2016}, ultracold gases~\cite{garay2000,garay2002,bs,uwe}, water and sound waves~\cite{unruh81,cg}. Here, we study the geo\-me\-tri\-za\-tion of soliton equation by considering a canonical field quantization in the classical background of the Sine-Gordon (SG) soliton metric.
Indeed, the $1+1$ dimensional Sine-Gordon (SG) equation 
\beq
\label{eq1}
\phi_{tt} - \phi_{xx} + m^2\sin(\phi)=0
\eeq
is a nonlinear model that exhibits a Riemannian surface with constant negative curvature.
%This model is interesting in quantum gra\-vi\-ty since it is possible to cast the Einstein-Hilbert action of a two dimensional dilaton gra\-vi\-ty theory into the SG one \cite{Gegenberg97}. 

In this frame, the SG equation can be considered the AKNS counterpart of a two dimensional gra\-vi\-ta\-tio\-nal theory. Two dimensional theories of gra\-vi\-ty are useful models to  understand the quantum properties of higher-dimensional gra\-vi\-ty. These theories capture essential features of higher-dimensional counterparts, and in particular have black hole solutions and Haw\-king radiation~\cite{giddings93,mandal91,witten91,callan92}. The link between the 1+1 dimensional gra\-vi\-ty and the SG model introduces further simplifications since the quantum properties of this equation have been largely studied~\cite{Faddeev77,zamolodchikov79}. As we shall recall in the next section, the integrability condition of SG equation determines a metric, with a coordinate singularity, which defines  an EH. In particular 1+1 dimensional BHs can be realized as solitons of the SG equation~\cite{Gegenberg97} and it has been shown with a one loop perturbative computation that this BH emits thermal radiation~\cite{vaz95,kim95}.

%Recently, Haw\-king radiation analogues from  solitons were considered in a huge variety of physical contexts, including light \cite{lp,s,bl,tc,villari2016}, ultracold gases \cite{garay2000,garay2002,bs,uwe}, water and sound waves \cite{unruh81,cg}. 
In this paper, we show that SG soliton emits thermal particles with a specific Haw\-king temperature, finding the way the temperature changes with the velocity of the SG-BH. Afterward, we perform two different kinds of quantization, one for a massless scalar field and another for the soliton itself, and obtain their Haw\-king emission spectra. In both cases, we discover that an observer on the soliton tail detects a thermal radiation with a temperature directly proportional to the soliton speed. Furthermore, we analyze the temperature detected by an observer at rest by adding a Doppler effect.

Our paper is organized as follows: in  sec. II we review the geometrization of the SG model;
we show the connection between a soliton solution of an AKNS system and a metric on a two dimensional surface. 
In  sec.  III we study the BH metric induced by the SG equation and introduce suitable coordinate systems for the field quantization. In section IV we quantize  massless scalar fields on the soliton background.  In section V we quantize the SG soliton following the Faddeev semiclassical quantization~\cite{Faddeev77}, and show that the sine-gordon BH evaporates. Conclusions are drawn in section VI.
A short appendix furnishes a minimal mathematical background to forms and curvature.

\section{Sine-Gordon geometrization}
We start reviewing the way integrable nonlinear equations generates surfaces with constant negative curvature~\cite{Bullough97}. 
By considering the SG equation defined in Eq.~(\ref{eq1}), we perform the coordinate transformation
\beq
\label{trasf1}
\chi=\frac{m}{2}(x+t),\;\;\theta=\frac{m}{2}(x-t)
\eeq
and get
\beq
\label{sg2}
\phi_{\chi\theta}=\sin\phi.
\eeq

As originally stated by Ablowitz, Kaup, Newel and Segur \cite{ablowitz73a}, for Eq. (\ref{sg2}) the following system defines the scattering problem  
\beq
\begin{cases}
\bold V_{\chi} = \hat L \bold V	\\
\bold V_{\theta} = \hat M \bold V
\end{cases},
\eeq
where $\hat L$ and $\hat M$ are $2 \times 2$ matrices, defining the Lax pair for Eq. (\ref{sg2}). $\bold V$ is a vector. This system corresponds to the integrable Pfaffian system \cite{flanders63} (see appendix for an introduction to forms and surfaces)
\beq
\D \bold V = \hat \Omega \bold V, \quad \bold V=\left(\begin{array}{c}V_1 \\V_2\end{array}\right), 
\eeq
where $\hat \Omega$ is a traceless matrix 
\beq
\hat \Omega = \hat L \D \chi + \hat M \D \theta = \left(\begin{array}{cc}\omega_1 & \omega_2 \\\omega_3 & -\omega_1\end{array}\right),
\eeq
with the matrix e\-le\-ments $\omega_i$ given by \cite{Bullough97} 
\beq
\begin{aligned}
\omega_1 &= -\frac{1}{2} \lambda \D \chi - \frac{1}{2\lambda}\cos(\phi) \D \theta,\\
\omega_2 &= -\frac{1}{2} \phi_{\chi} \D \chi - \frac{1}{2 \lambda} \sin(\phi) \D \theta,\\
\omega_3 &=  \frac{1}{2} \phi_{\chi} \D \chi - \frac{1}{2 \lambda} \sin(\phi) \D \theta,\\
\end{aligned}
\eeq
where $\lambda$ is the spectral parameter of the SG scattering problem \cite{ablowitz73a}. 
Following \cite{sasaky79}, the arclength of the induced Riemannian surface is written in terms of the matrix e\-le\-ments $\omega_i$ as follows \cite{flanders63, Bullough97}
\beq
\begin{aligned}
\D s^2 = (\omega_2 + \omega_3)^2 + (2\omega_1)^2=\\=\lambda^2\D\chi^2+2\cos(\phi)\D\chi\D\theta+\frac{1}{\lambda^2}\D\theta^2.
\end{aligned}
\label{metric1}
\eeq
Eq.~(\ref{metric1}) defines the constant negative curvature metric induced by the ISM associated to the SG equation~(\ref{sg2}).
By changing the coordinates set as in the following, we write the first fundamental form $\D s^2$ as~\cite{Bullough97}
\beq
\label{eq2}
\D s^2 =\sin^2\Bigl(\frac{\phi}2 \Bigl)\D \tau^2+\cos^2\Bigl(\frac{\phi}2 \Bigl)\D \xi^2,
\eeq
which results to be associated with a SG equation of the form
\beq
\label{eq:SG}
\phi_{\xi\xi}-\phi_{\tau\tau}=\sin(\phi),
\eeq
where
\beq
\label{trasf2}
\left\{\begin{array}{c}\xi=\lambda\chi+\lambda^{-1}\theta\\\tau=\lambda\chi-\lambda^{-1}\theta\end{array}\right..
\eeq

Thus the metric tensor is
\beq
\hat g = \left(\begin{array}{cc} g_{\tau\tau} & g_{\xi\tau} \\ g_{\tau\xi} & g_{\xi\xi}\end{array}\right)=\left(\begin{array}{cc}\sin^2 \frac{\phi}2 & 0 \\0 & \cos^2 \frac{\phi}2\end{array}\right).
\eeq
However, $\D s^2$ in Eq. (\ref{eq2}) is not Lorentz invariant and it does not lead to a Schwarzschild-like metric. Following \cite{Gegenberg97}, in order to obtain a Minkowski-like metric, we perform a Wick rotation $\tau\rightarrow i\tau$ and obtain the elliptic SG (ESG) equation:
\beq
\label{eq:ESG}
\phi_{\xi\xi}+\phi_{\tau\tau}=\sin(\phi),
\eeq
whose corresponding metric is
\beq
\label{metrica}
\D s^2 =-\sin^2\Bigl(\frac{\phi}2 \Bigl)\D \tau^2+\cos^2\Bigl(\frac{\phi}2 \Bigl)\D \xi^2.
\eeq
%%%%%%%%%%%%%%%%%%%%%%%%%%%%%%%%%%%%%%%%%%%%%%%%%%%%%%%%%%%%%%%%%%%%%
\section{\label{sec:sgbh} The Sine-Gordon soliton black hole}
%%%%%%%%%%%%%%%%%%%%%%%%%%%%%%%%%%%%%%%%%%%%%%%%%%%%%%%%%%%%%%%%%%%%%

We show that the one-soliton solution of the ESG equation determines a BH metric.

\begin{figure}[H]
\begin{center}
\includegraphics[width=1 \columnwidth]{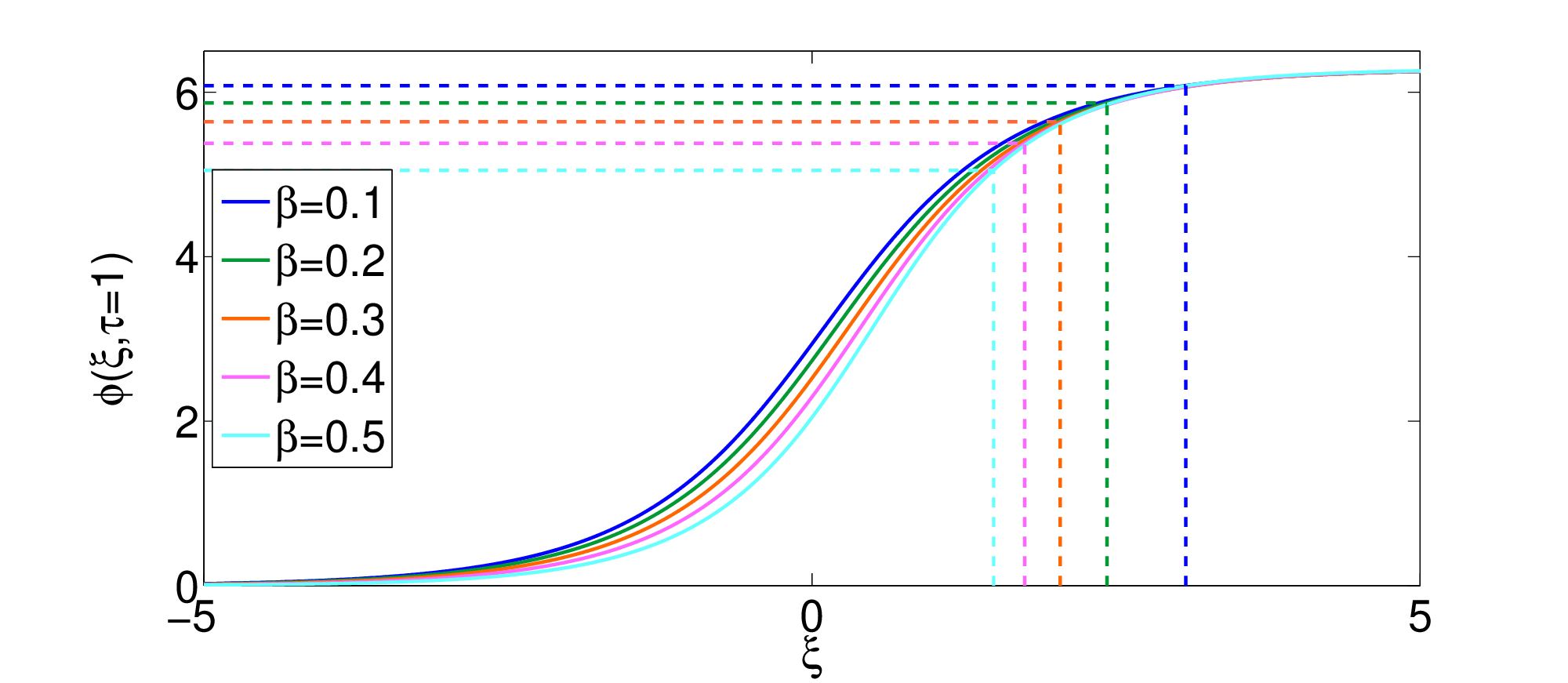}
\caption{(Color online) The sine-Gordon soliton at fixed time $\tau=1$, varying the velocity $\beta_s$. The positions of the EHs $\rho_g=m\gamma(\xi_g-\beta_s \tau)$ are in dashed lines. }
\label{fig_soliton}
\end{center}
\end{figure}

\begin{figure}[H]
\begin{center}
\includegraphics[width=1 \columnwidth]{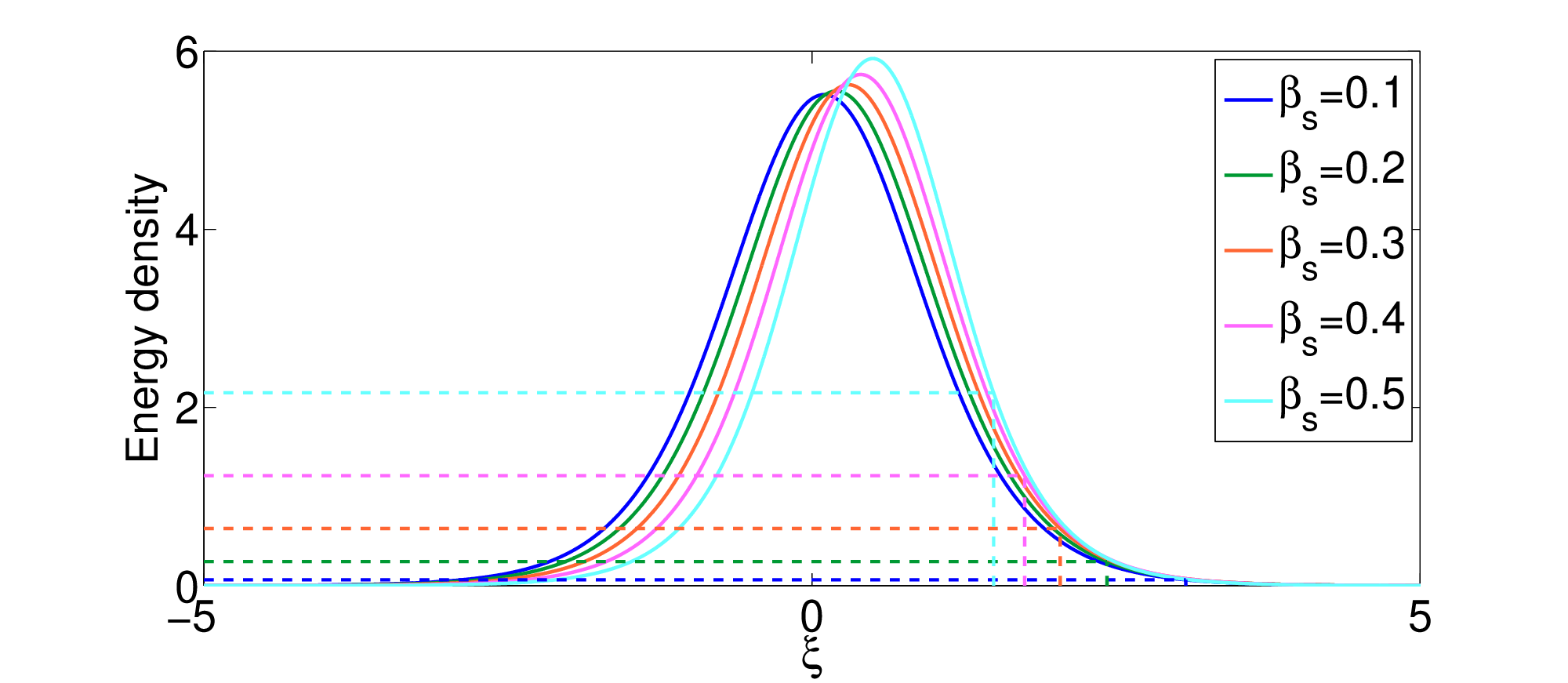}
\caption{(Color online) The soliton energy density at various velocities $\beta_s$, at fixed time $\tau=1$. The positions of the EHs $\rho_g=m\gamma(\xi_g-\beta_s \tau)$ are in dashed lines.}
\label{fig_energy}
\end{center}
\end{figure}

The well known forward-propagating one-soliton solution of the Eq. (\ref{eq:SG}) is
\beq
\label{solitone}
\phi(\xi,\tau)=4\arctan\left\{\exp\left[\gamma\left(\xi-\beta_s\tau\right)\right]\right\},
\eeq
with $\gamma=(1-\beta_s^2)^{-1/2}$ and $0<\beta_s<1$ the soliton velocity \cite{ablowitz74}. The backward-propagating one-soliton solution gives the same treatise with $-1<\beta_s<0$, by substituting $\beta_s$ in $-\beta_s$ in what follows. For this reason, we can choose solution (\ref{solitone}) without loss of generality.
Eq. (\ref{solitone}) is also solution of Eq. (\ref{eq:ESG}) with
\beq
\label{eq:gamma}
\gamma=(1+\beta_s^2)^{-1/2}.
\eeq
We adopt Eq. (\ref{eq:gamma}) hereafter.
Substituting Eq. (\ref{solitone}) in Eq. (\ref{metrica}), we have 
\beq
\label{eq4}
ds^2=ds^2_{1sol} = -\sech^2(\rho) d\tau^2 + \tanh^2(\rho) d\xi^2 ,
\eeq
with $\rho = \gamma(\xi-\beta_s\tau)$.  Following \cite{Gegenberg97}, we adopt various coordinate transformations: first from $(\tau,\xi)$ to $(\mathcal{T},\rho)$, with $\rho$ as defined above and 
\beq \label{eq5}
\mathcal{T} = \tau- \frac{1}{\beta_s}\{ \tanh^{-1}[\gamma^{-1} \tanh (\rho) ] - \gamma^{-1}\rho\}.
 \eeq
Next, we transform $(\mathcal{T},\rho)$ to $(\mathcal{T},r)$ by 
\beq \label{eq6}
r = \frac{1}{\gamma} \sech(\rho).
\eeq
The result of the transformation is the line e\-le\-ment 
\beq \label{eq7}
ds^2 = (\beta_s^2-r^2) d\mathcal{T}^2 - (\beta_s^2-r^2)^{-1} dr^2.
\eeq
Eq. (\ref{eq7}) is the metric of a 1+1 dimensional BH with EH at $r_g:= \beta_s$. Figures \ref{fig_soliton} and \ref{fig_energy} show the EH positions \\$\rho_g=\arcsech(\gamma r_g)$ on the soliton profiles and e\-ner\-gy densities $\mathcal{E}$, respectively for different velocities $\beta_s$. The e\-ner\-gy density, at fixed $t$, is defined as follows \cite{Faddeev77}:
$$
\mathcal{E}=\frac 1 2(\partial_{\xi}\phi_s)^2+[1-cos(\phi_s)].
$$
It is now convenient to introduce two new sets of coor\-di\-nates: the modified  \emph{Regge-Wheeler coor\-di\-nate}, that we call the \emph{slug coor\-di\-nate} in analogy with the \emph{tortoise coor\-di\-nate}, as usually reported \cite{hawking74,mwi}, and the \emph{Kruskal-Szekeres coor\-di\-nates}.\\
We get the slug coordinate $r^*(r)$ according to
\beq \label{eq8}
dr^* =(\beta_s^2-r^2)^{-1} dr,
\eeq
so that 
\beq \label{eq9}
r^*(r)= \frac{1}{\beta_s} \tanh^{-1} \left(\frac{r}{\beta_s}\right)=\frac{1}{2\beta_s}\ln\left(\frac{\beta_s+r}{\beta_s-r}\right).
\eeq
Eq. (\ref{eq7}) then becomes 
\beq \label{eq10}
ds^2 = [\beta_s^2-r^2(r^*)][d\mathcal{T}^2-(dr^*)^2] .
\eeq
The slug coordinate is singular at $r=\beta_s$ and it is defined on the exterior of the BH when  $\rho \to \pm \infty$  and $r \to 0$.  In fact, as $r$ approaches $\beta_s$, $r^*$ goes to $+\infty$, while far away from the BH $r^* \to 0$ as $r \to 0$.\\
Introducing the slug lightcone coor\-di\-nates
\beq \label{eq11}
\tilde u = \mathcal{T}-r^*, \quad \tilde v = \mathcal{T} + r^*,
\eeq
we write Eq. (\ref{eq7}) as
\beq \label{eq12}
ds^2 = [\beta_s^2-r^2(\tilde u,\tilde v)]d\tilde u d\tilde v .
\eeq
The slug lightcone coor\-di\-nates are singular and they span only the exterior of the black hole. To describe the entire spacetime, we need another coordinate system. In order to be consistent with literature, we refer to them as the Kruskal-Szekeres (KS) coordinates . From Eq. (\ref{eq9}) and Eq. (\ref{eq11}) it follows that  
\beq \label{eq13}
 \beta_s^2-r^2 = (\beta_s+r)^2 \exp[\beta_s(\tilde u - \tilde v)].
 \eeq
 The BH metric thus becomes 
 \beq \label{eq14}
 ds^2 = [\beta_s+r(\tilde u,\tilde v)]^2 e^{\beta_s(\tilde u - \tilde v)}d\tilde u d\tilde v.
 \eeq 
 In the KS lightcone coordinates, defined as 
 \beq \label{eq15}
 u = \frac{e^{\beta_s\tilde u}}{\beta_s}\text{,} \quad v = -\frac{e^{-\beta_s\tilde v}}{\beta_s},
 \eeq
 Eq. (\ref{eq14}) takes  the form 
 \beq \label{eq16}
 ds^2 = [\beta_s+r(\tilde u,\tilde v)]^2 dudv,
 \eeq
and it is  regular at $r=\beta_s$. The singularity occurring in the ESG-soliton metric  is, as the Schwarzschild one, a coordinate singularity, which can be removed by a coordinate transformation. The KS coordinates, indeed, span the entire spacetime.

%%%%%%%%%%%%%%%%%%%%%%%%%%%%%%%%%%%%%%%%%%%%
\section{Massless scalar field quantization}
%%%%%%%%%%%%%%%%%%%%%%%%%%%%%%%%%%%%%%%%%%%%

We consider a field quantization on the classical soliton background metric. We first analyze a massless scalar field with the action 
\beq \label{eq17}
S[\phi] = \frac{1}2 \int g^{\mu \nu} \partial_\mu \phi \partial_\nu \phi \sqrt{-g} \, d^2\underline{x},
\eeq
where $g^{\mu \nu}$ represents the inverse of a general metric tensor $g_{\mu \nu}$, $g$ is the determinant of $g_{\mu \nu}$ and $\underline{x}=(x^0,x^1)$.
The action in Eq. (\ref{eq17}) is conformally invariant, and in terms of lightcone slug coor\-di\-nates and lightcone KS coor\-di\-nates (\ref{eq16}) it reads 
\beq \label{eq18}
\begin{aligned}
S[\phi] = \int \partial_{\tilde u}\phi \partial_{\tilde v}\phi \, d\tilde u d\tilde v, \\
S[\phi] = \int \partial_{u}\phi \partial_{v}\phi \, du dv.
\end{aligned}
\eeq
We write the solution of the scalar field equation in terms of the lightcone slug coordinates
\beq \label{eq19}
\phi = \tilde A(\tilde u) + \tilde B (\tilde v),
\eeq
and in the lightcone KS coordinate as
\beq \label{eq20}
\phi = A(u) + B(v),
\eeq
where $A$, $\tilde A$ and $B$, $\tilde B$ are arbitrary smooth functions. 
In correspondance of the tail of the soliton, i.e., far away from the EH, 
the mode expansion of the field is  
\beq \label{eq21}
\hatF= \int_0^{\infty} \frac{d\Omega}{2\sqrt{\pi \Omega}} \bigl[e^{-i \Omega \tilde u} \bm +  e^{+i \Omega \tilde u} \bp \bigl] + \text{left moving}.
\eeq
In Eq. (\ref{eq21}) the \emph{left moving} part is given by the terms weighted by $e^{\pm i\Omega \tilde v}$ in the mode expansion.  
The vacuum state $\k{0_B}$, defined by $\bm \k{0_B} =  0$, is the \emph{Boulware va\-cuum} (BV) and does not contain particles for an observer located far from the EH. 
However, as the slug coordinate is singular at horizon, the BV is also singular at the EH. 

\begin{figure}
\begin{center}
\includegraphics[width=1 \columnwidth]{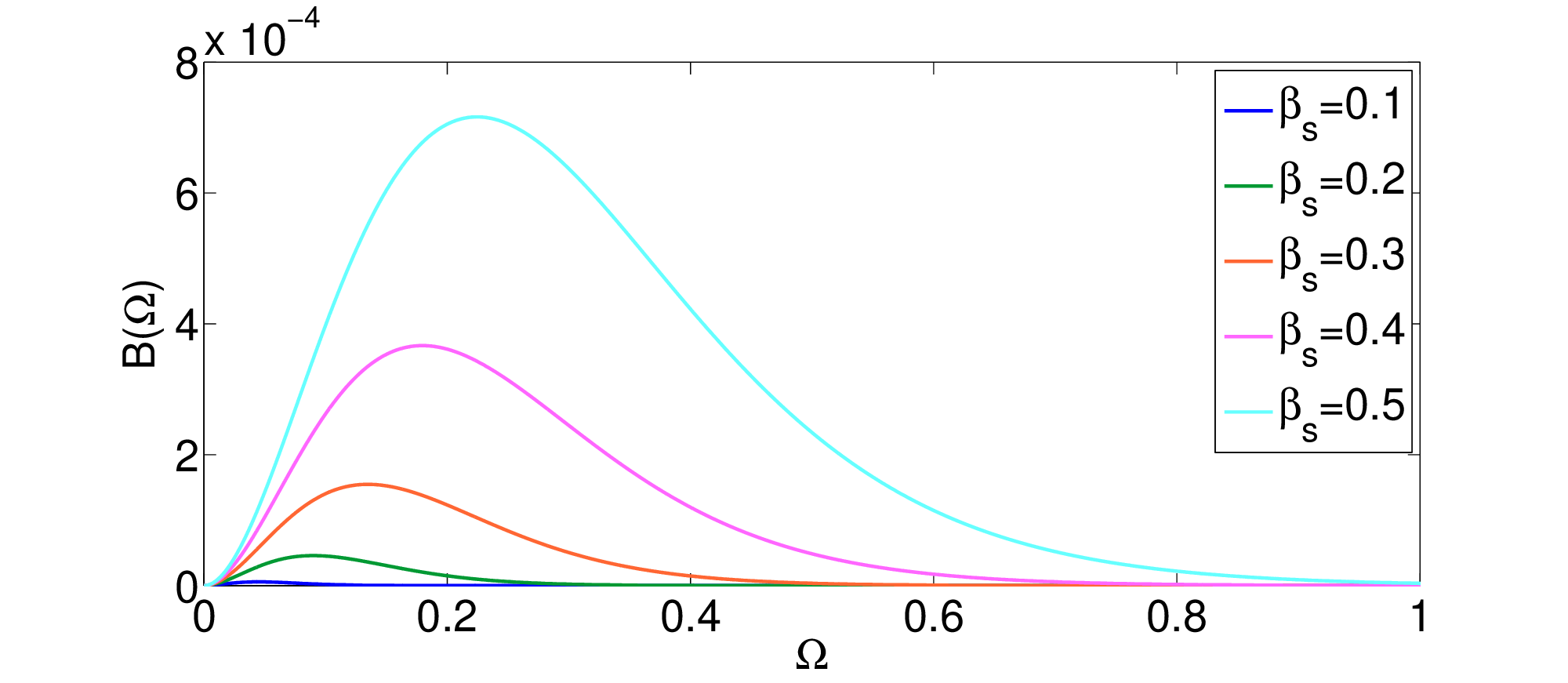}
\caption{(Color online) Spectral radiance for massless fields varying $\beta_s$.}
\label{fig_massless}
\end{center}
\end{figure}

To obtain a vacuum state defined over the entire spacetime, we expand the field operator in terms of the KS lightcone coordinates
\beq \label{eq22}
\hatF = \int_0^{\infty} \frac{d\omega}{2\sqrt{\pi \omega}} \bigl[e^{-i\omega u} \am + e^{i\omega u} \ap \bigl] + \text{left moving}.
\eeq
The creation and annihilation operators $\hat a^{\pm}_\omega$ determine the \emph{Kruskal vacuum} (KV) state $\am \k{0_K} =0$. The KV is regular on the horizon and corresponds to true physical vacuum in the presence of the BH.\\
For a remote observer the KV contains particles. To determine their number density, we follow the original calculations of Haw\-king and Unruh with the only dif\-fe\-rence in the definition of the KS coordinates (see chapters 8 and 9 of \cite{mwi} for details). \\
We find that the remote observer moving with the soliton tail sees particles with the thermal spectrum 
\beq \label{eq23}
\braket{\hat N_{\Omega}}=  \br{0_K }\bp \bm \k {0_K} = \left[\exp  \left(\frac{2\pi \Omega}{\beta_s}\right) -1 \right]^{-1} \delta(0).
\eeq
If we consider a finite volume quantization we can put $V=\delta(0)$ \cite{mwi} and we obtain the number density
\beq
n_{\Omega} =  \left[\exp \left(\frac{2\pi \Omega}{\beta_s}\right) -1 \right]^{-1},
\eeq
corresponding to the temperature 
\beq \label{eq24}
T_H = \frac{\beta_s}{2\pi} .
\eeq
In Fig. \ref{fig_massless}, we show the radiance $B(\Omega)=\Omega^3 n_{\Omega}$. We observe that for a static soliton ($\beta_s = 0$) we get $T_H=0$. This result may appear in contradiction  with the Haw\-king original work, where he considered the emission from a static BH. However the result in eq. (\ref{eq24}) is coherent with the structure of the metric induced by the SG equation, where the singularity occurs for $r = r_g = \beta_s$ and no emission can be observable for $r_g =0$. This dependence of the Haw\-king radiations on the translation velocity is peculiar of soliton dynamics~\cite{mp} and it is related to the structure of the spectral parameter in the IST~\cite{ablowitz73,ablowitz73a}.

%%%%%%%%%%%%%%%%%%%%%%%%%%%%%%%%%%%%%%%%%%%%%%%%%%%%%%%%
\subsection{Haw\-king temperature in the laboratory frame}
%%%%%%%%%%%%%%%%%%%%%%%%%%%%%%%%%%%%%%%%%%%%%%%%%%%%%%%%

Unlike the Schwarzschild BH, the ESG soliton is not static, but translates with velocity $\beta_s$. The frequency $\Omega$ seen by an observer at rest with respect to the soliton contains a Doppler shift. 
Letting $\Omega_s$ be the frequency emitted by the soliton in (\ref{eq23}), the frequency measured by an observed moving with velocity $-\beta_s$ with respect to the soliton, and located at an angle $\theta_s$ with respect to the soliton direction is 
\beq \label{eq25}
\Omega_o = \frac{1-\beta_s \cos \theta_s}{\sqrt{1-\beta_s^2}} \Omega_s.
\eeq
In the collinear case $\theta_s=0$, and we have 
\beq \label{eq27}
\frac{\Omega_o}{\Omega_s} = \sqrt{\frac{1-\beta_s}{1+\beta_s}} .
\eeq  
The corresponding  Haw\-king temperature is (for small $\beta_s$)
\beq \label{eq28}
T_H = \frac{\beta_s}{2\pi} \sqrt{\frac{1-\beta_s}{1+\beta_s}} \simeq \frac{\beta_s}{2\pi} (1-\beta_s).
\eeq

This calculation also applies to a massive bosonic field, as the number density spectrum depends only on the statistics \cite{crispino2008}. In the case of a fermionic field the theo\-ry is similar, but the  number density spectrum follows the Fermi-Dirac statistics \cite{hawking74}.
%%%%%%%%%%%%%%%%%%%%%%%%%%%%%
\section{Soliton Quantization}
\begin{figure}[H]
\begin{center}
\includegraphics[width=1 \columnwidth]{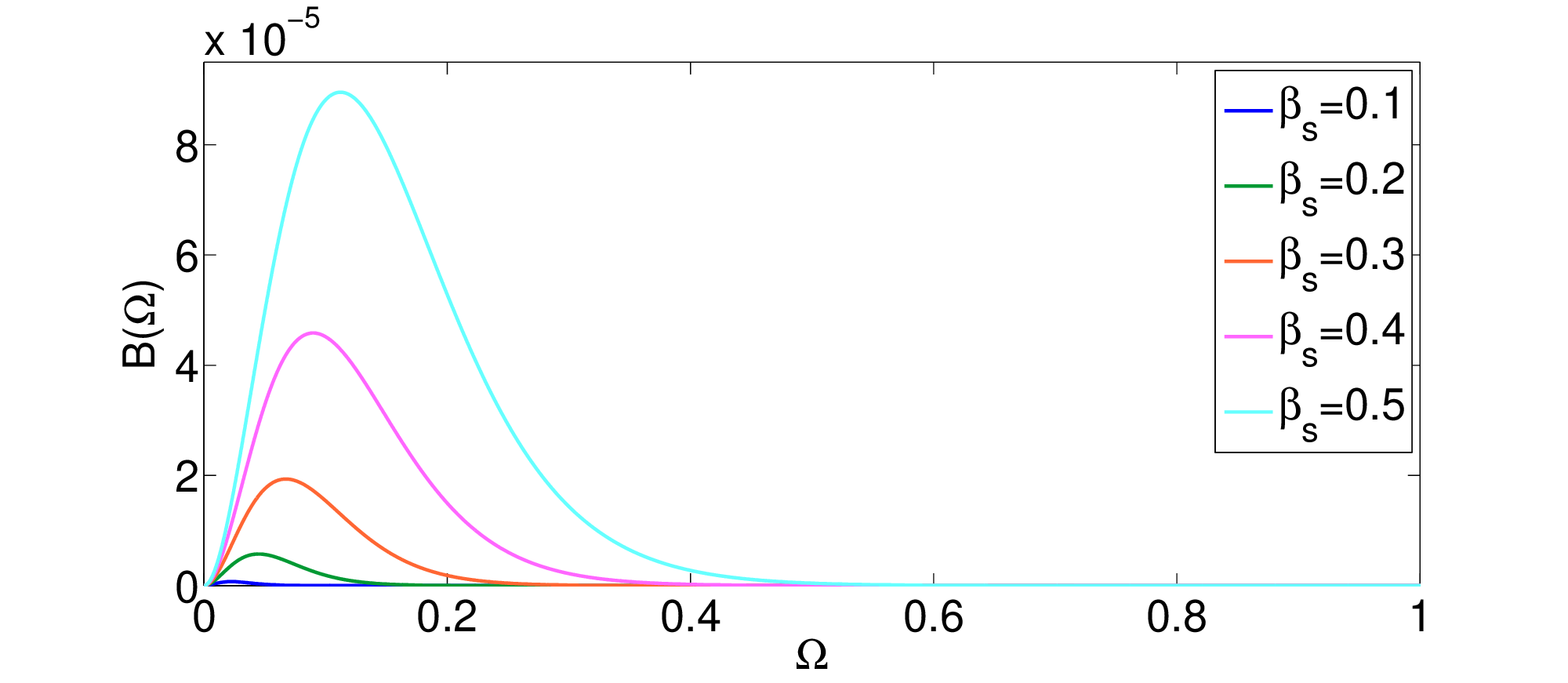}
\caption{(Color online) Spectral radiance for massive fields varying the soliton velocity.}
\label{fig_massive}
\end{center}
\end{figure}

%%%%%%%%%%%%%%%%%%%%%%%%%%%%%%
Previously we studied the BH evaporation following the works of Haw\-king and Unruh in \cite{hawking74,Unruh1976}. Now, we analyze a quantum perturbation of the BH metric given by the classical soliton solution of the ESG equation, and we obtain a BH evaporation without the interaction with a massless scalar field.
We start from
\beq \label{eq38}
\phi \simeq \phi_s+\phi_1,
\eeq
where $\phi_s$ is the classical solution in Eq. (\ref{solitone}) and $\phi_1$ represents a weak field perturbation.
We consider the conformally invariant action
\beq
\label{eq:sgaction}
S[\phi] = \int  \,  \left[\frac 1 2 g^{\mu\nu}\partial_\mu \partial_\nu \phi+cos(\phi)\right] \sqrt{-g}  \, d^2x,
\eeq
which leads to a field equation
\beq
\label{eq:covariantsg}
g^{\mu\nu} \partial_\mu \partial_\nu \phi +  \sin(\phi) =0.
\eeq
%with $\square=g^{\mu\nu}\partial_{\mu}\partial_{\nu}$ . 
The solutions of Eqs. (\ref{eq:SG},\ref{eq:ESG}) differ for a Wick rotation. In other words, one passes from the SG soliton to the ESG one by the transformation
\beq
\label{eq:wick}
\tau\rightarrow i\tau,\;\;\beta_s\rightarrow -i\beta_s.
\eeq
We perform the inverse Wick rotation, i.e., $\tau\rightarrow -i\tau$, \\$\beta_s\rightarrow i\beta_s$, passing from the ESG to the SG, and substitute Eq. (\ref{eq38}) into Eq. (\ref{eq:covariantsg}), hence we obtain
\beq
g^{\mu\nu} \partial_\mu \partial_\nu \phi_1 + \cos(\phi_s)\phi_1=0,
\eeq
where we neglect terms $O(\phi_1^2)$. 
This equation expresses the interaction between a massive particle and the gra\-vi\-ta\-tio\-nal field, because the weak quantum field $\phi_1$ obeys a \emph{generalized  Klein-Gordon} equation with squared mass $\cos(\phi_s)$ depending on the soliton, and thus on the metric. Recalling Eq.~(\ref{solitone}), we have 
\beq
g^{\mu\nu} \partial_\mu \partial_\nu \phi_1+\cos\{4\arctan[\exp(\rho)]\}\phi_1=0.
\eeq
For an observer located on the tail of the soliton \\($\rho \to \infty$), the field equation reduces to
\beq
g^{\mu\nu} \partial_\mu \partial_\nu  \phi_1 + \phi_1 =0,
\eeq
while for an observer on the horizon ($\rho \to \rho_g$), we have
\beq
g^{\mu\nu} \partial_\mu \partial_\nu \phi_1 + F(\rho) \phi_1 =0,
\eeq
with $F(\rho)$ given by 
\beq
 F(\rho)|_{\rho\sim\rho_g} \simeq 1+ \frac{5}{2}\gamma^2\beta_s^2 - 5\gamma^2\beta_s^2\sqrt{1-\gamma^2v^2}(\rho-\rho_g).
\label{eqcum}
\eeq
Eq. (\ref{eqcum}) truncated at the order zero in $\rho-\rho_g$, i.e., exactly on the horizon, leads to
\beq
F \simeq 1+ \frac{5}{2}\gamma^2\beta_s^2.
\eeq
Due to the inverse Wick rotation, even if the action is conformally invariant, the quantization is not straightforward. We need to adapt both the slug and the KS lightcone coordinates in Eqs. (\ref{eq11},\ref{eq15}) to the rotated sy\-stem. We obtain
\beq \label{eq41}
\begin{aligned}
r^*(r) &= \int_0^r \frac{dr'}{\beta_s^2 + r'^2} = \frac{i}{2\beta_s}\ln{\Biggl(\frac{i\beta_s + r}{i\beta_s - r}\Biggl)}, \\
\tilde u &= \mathcal T - ir^*, \quad \tilde v = \mathcal T + ir^*,\\
u &= -\frac{e^{-\beta_s \tilde u}}{\beta_s}, \quad v = \frac{e^{\beta_s \tilde v}}{\beta_s}.
\end{aligned}
\eeq
Since the action (\ref{eq:sgaction}) is conformally invariant, we thus write the field equation as follows
\beq
\begin{aligned}
&\paut\pavt \phi_1 + \phi_1 =0 \,\, \, \text{slug lightcone}, \\
&\pau \pav \phi_1 + F   \phi_1 =0 \,\,\, \text{ K-S lightcone},
\label{eq45}
\end{aligned}
\eeq
Eqs. (\ref{eq45}) have exponential solution
\beq
\begin{aligned}
\phi_1 \propto &  e^{i(K-\Omega_K)\tilde u-i(K+\Omega_K)\tilde v},\\
\phi_1 \propto &  e^{i(k-\omega_k) u -i(k+\omega_k)v},
\end{aligned}
\eeq
with the following dispersion relations, 
\beq
\begin{aligned}
\Omega_K & = \sqrt{K^2 + 1},\\
\omega_k &= \sqrt{k^2 + F^2}.
\end{aligned}
\eeq
From now on, we omit the $K$ and $k$ indices. We write the quantum fields as follows 
\begin{widetext}
\beq \label{eq42}
\begin{aligned}
\hatF_0 &= \frac{1}{2\pi} \int_0^\infty \frac{d\Omega}{\sqrt \Omega} \, [\bm e^{i(K-\Omega)\tilde u-i(k+\Omega)\tilde v} + \bp  e^{-i(K-\Omega)\tilde u+i(K-\Omega)\tilde v}]=\\
 &= \frac{1}{2\pi}  \int_0^\infty \frac{d\omega}{\sqrt \omega}\, [\am e^{i(k-\omega) u -i(k+\omega)v} +  \ap  e^{-i(k-\omega)u+i(k+\omega)v}],
\end{aligned}
\eeq
\end{widetext}
where, as in the non interacting case, the annihilation operators $\bm$ and $\am$ define the Boulware vacuum $\k{0_B}$ and the Kruskal vacuum $\k{0_K}$, respectively.
The operators $\hat a_\omega^{\pm}$ and   $\hat b_\Omega^{\pm}$ are related by the Bogolyubov transformations
\beq \label{eq43}
\bm = \int_0^\infty d\omega \, (\a \am - \b \ap).
\eeq
By substituting this in Eq.~(\ref{eq42}), we find
\begin{widetext}
\beq \label{eq44}
\frac{1}{\sqrt{\omega}} \int_{-\infty }^{\infty} d\tilde u d\tilde v\, e^{i[\Omega(\tilde u + \tilde v) - K(\tilde u - \tilde v) } e^{-i[\omega(u+v) +k(u-v)]}=\int_0^\infty \frac{d\Omega'}{\sqrt{\Omega'}} \, \alpha_{\Omega' \omega} [2\pi \delta(\Omega-\Omega')]^2,
\eeq
\end{widetext}
hence we obtain 
\beq
\alpha_{\omega \Omega} = \frac{1}{2\pi V} \sqrt{\frac{\Omega}{\omega}} \int d\tilde u d\tilde v e^{iu(k-\omega)-iv(k+\omega)} e^{-i\tilde u(K-\Omega) +iv(k+\Omega)}.
\eeq
Seemingly for $\b$, we have 
\beq \label{eq:beta}
\beta_{\omega \Omega} =  -\frac{1}{2\pi V} \sqrt{\frac{\Omega}{ \omega}} \int d\tilde u d\tilde v e^{-iu(k-\omega)+iv(k+\omega)} e^{-i\tilde u(K-\Omega) +iv(k+\Omega)}.
\eeq
Using now the KS coordinate (\ref{eq41}), after lengthy but straightforward calculations, we find 
\begin{widetext}
\beq
\begin{aligned}
\a &= \frac{1}{2\pi V } \frac{\Omega}{\omega} e^{\pi \Omega / \beta_s}  e^{iF(\Omega,K,\omega,k)} \Gamma\Bigl[i\frac{\Omega+K}{\beta_s}\Bigl]  \Gamma\Bigl[i\frac{\Omega-k}{\beta_s}\Bigl], \\
\b &= \frac{1}{2 \pi V } \frac{\Omega}{\omega} e^{-\pi \Omega/ \beta_s}  e^{iG(\Omega,K,\omega,k)} \Gamma\Bigl[i\frac{\Omega+K}{\beta_s}\Bigl]  \Gamma\Bigl[i\frac{\Omega-k}{\beta_s}\Bigl]. 
\end{aligned}
\eeq
\end{widetext}
It follows that $\a$  and  $\b$  obey the useful relation
\beq
|\a|^2 = e^{4\pi\Omega/\beta_s} |\b|^2.
\eeq
Therefore we can compute the expectation value of the $b$-particle number operator $\hat N_{\Omega} = \bp\bm$ in the Kruskal vacuum \cite{mwi}, and obtain the number density 
\beq
n_{\Omega}= \Bigl[\exp \Bigl(\frac{2 \Omega }{T_H}\Bigr) -1 \Bigl]^{-1}. 
\eeq
This corresponds to an emitted radiation with twice the frequency with respect to the simple massless case, of which spectral radiance  $B(\Omega)=\Omega^3 n_{\Omega}$ is reported in figure~\ref{fig_massive}. We observe that the Haw\-king temperature is equal to Eq.~(\ref{eq24}) for the massless scalar field. This is expected since the surface gra\-vi\-ty is the same. For a moving observer with respect to the soliton the Haw\-king temperature, for small $\beta_s$, reads
\beq
T_H = \frac{\beta_s}{2\pi} \sqrt{\frac{1-\beta_s}{1+\beta_s}} \simeq \frac{\beta_s}{2\pi} (1-\beta_s).
\label{eqfinal}
\eeq
Eq. (\ref{eqfinal}) provides the Haw\-king temperature of soliton evaporation in this toy model. 
\section{Conclusions}
We adopted the geometrization of the ESG model and reported on the connection between the one-soliton solution of the 1+1-dimensional elliptic sine-Gordon equation and a metric with a Schwarzschild-like coordinate singularity.  We determined the BH metric and, by suitable coordinate systems, we eliminated the singularity and obtained a regular metric on the EH. We quantized a massless scalar field and found the thermal ra\-dia\-tion detected by an observer far away on the BH exterior. We obtained that the temperature is proportional to the soliton velocity. We analyzed the temperature detected by an observer in the laboratory frame, by a Doppler effect.
We studied also the quantum soliton evaporation, and found the corresponding spectrum.

Our analysis allows to predict the Haw\-king radiation for a moving 1+1 dimensional BH and shows that the velocity affects the temperature and the corresponding emitted thermal spectrum.
In a BH collisional process one can hence expect a frequency shift of the emitted photon concomitant with the variation of spiraling velocity of the BHs.
The resulting chirp of the emitted photons may have a clear and detectable signature in the electromagnetic spectrum. Analogues of these processes may be eventually simulated in the long-range interactions between opti\-cal solitons pairs recently observed over astronomical distances \cite{JE2013}, or similar opti\-cal experiments \cite{Bek2015,Faccio2016}.

Our results may be extended to any metric induced by AKNS systems, hence to many different physical models to conceive experimentally realizable analogues for studying Hawking evaporation of moving black holes.

\section*{Acknowledgments}
We are pleased to acknowledge Prof. Fabio Biancalana for invaluable discussions and for a critical rea\-ding of the manuscript.
We also acknowledge I. M. Deen for technical support with the computational resources.
C.C. and G.M. acknowledge support from the Templeton Foundation (grant number 58277), the  H2020 QuantERA project QUOMPLEX (project ID 731473) and PRIN project NEMO (ref. 2015KEZNYM).

\section*{Appendix: minimal introduction to forms, Pfaff problems and curvature}
A 1-form $\Omega=X \D x+T \D t$ is a combination of the differentials $\D x$ and $\D t$, which have to be retained as e\-le\-ments of a basis. $X$ and $T$ can be matrices with the same size, or also operators.
A 2-form is a combination of the symbols (``exterior products'') $\D x\wedge \D t$ and \\$\D t\wedge \D x=-\D x\wedge \D t$.
One can obtain a 2-form from a 1-form by the differential operator $\D $:
\begin{equation}
\D \Omega= \frac{\partial X}{\partial t} \D t\wedge \D x+ \frac{\partial T}{\partial x} \D x\wedge \D t=\left(-\frac{\partial X}{\partial t}+\frac{\partial T}{\partial x}\right)\D x \wedge \D t\text{,}
\end{equation}
which can be kept in mind by letting \\$\D x \wedge\D x=\D t \wedge\D t=0$,
so that terms like $\partial_x X$ and $\partial_t T$ do not appear in $\D \Omega$.
\\One can also obtain a 2-form by the exterior product $\Omega\wedge\Omega$ again by  $\D x \wedge\D x=\D t \wedge\D t=0$
\begin{equation}
\Omega\wedge\Omega= X T \D x \wedge \D t + T X \D t \wedge \D x=[X,T]\D x\wedge \D t
\end{equation}
with $[X,T]$ the commutator.
\\By using forms, the AKNS integrability condition 
\begin{equation}
\frac{\partial X}{\partial t}-\frac{\partial T}{\partial x}+[X,T]=0\text{,}
\end{equation}
reads as
\begin{equation}
\D \Omega -\Omega\wedge\Omega=0.
\label{int1}
\end{equation}
For some authors, using forms has the advantage of a more compact notation as the explicit coordinates $x$ and $t$ do not appear in (\ref{int1}).
Eq. (\ref{int1}) is referred to a Pfaffian integrability condition, or Pfaff problem.

Forms are directly connected to the curvature of surfaces.
If one considers a surface, and a local point vector ${\bf P}$ on the surface, let ${\bf e}_1$ and ${\bf e}_2$ the orthogonal tangent vectors.
For infinitesimal motion on the surface $\D {\bf P}$
\begin{equation}
\D {\bf P}=\sigma^1 {\bf e}_1+\sigma^2 {\bf e}_2\text{,}
\end{equation}
where $\sigma^1$ and $\sigma^2$ contain the differentials of the adopted coordinates and are hence 1-form.  $\sigma^1 \wedge \sigma^2$ is the e\-le\-mental area on the surface. 
When one moves of an amount $\D {\bf P}$, ${\bf e}_{1,2}$ changes of amounts $\D {\bf e}_{1,2}$. One considers a surface such that $\D {\bf e}_1=\omega {\bf e}_2$ and  $\D {\bf e}_2=-\omega {\bf e}_1$ where $\omega$ depends on the shape of the surface, contains the differentials of the coordinate systems, and is a 1-form named the {\it connection one form}. One finds the following equation
\begin{equation}
\D \omega=-K \sigma^1\wedge \sigma^2
\label{k1}
\end{equation}
where $K$ is the Gaussian curvature. $\omega$, $\sigma^1$ and $\sigma^2$ are one forms that fix all the properties of the surface.
In the particular case $K=-1$, one has from (\ref{k1})
\begin{equation}
\D \omega=\sigma^1\wedge \sigma^2.
\label{k11}
\end{equation}
By using (\ref{k11}) and considering the matrix 1-form \cite{sasaky79}
\beq
\Omega = \left(\begin{array}{cc} -\frac{1}{2}\sigma^2 & \frac{1}{2}(\omega +\sigma^1) \\\frac{1}{2}(-\omega+\sigma^1) & \frac{1}{2}\sigma^2 \end{array}\right),
\eeq
one finds the Pfaff system in Eq. (\ref{int1}).
In other words, considering the integrability condition (\ref{int1}), and re\-tai\-ning the e\-le\-ment of $\Omega$ as the forms of a two-dimensional surface, Eq. (\ref{int1}) implies that the surface has a constant negative curvature $K=-1$.
Hence integrability produces pseudospherical surfaces, i.e., surfaces of constant negative curvature.

%\bibliography{Bibliografia_SG}

\end{document}